# Anisotropic thermal expansion and thermomechanic properties of monolayer β-Te

Gang Liu,[⊥,†,*] Zhibin Gao,[§,ζ†] Jie Ren[§,*]

[⊥]School of Physics and Engineering, Henan University of Science and Technology, Luoyang 471023, China.

[§]Center for Phononics and Thermal Energy Science, China-EU Joint Center for Nanophononics, Shanghai Key Laboratory of Special Artificial Microstructure Materials and Technology, School of Physics Sciences and Engineering, Tongji University, Shanghai 200092, China

[ζ]Department of Physics, National University of Singapore, Singapore 117551, Republic of Singapore

## ABSTRACT

Recently, β-Te (atomically 2D tellurium) with rectangular crystal structure has been synthesized successfully on highly oriented pyrolytic graphite substrates by using molecular beam epitaxy (Zhu et al, Phys. Rev. Lett. **119**, 106101 (2017); Chen et al, Nanoscale **9**, 15945–15948 (2017).). It has been found possessing remarkable properties such as ultralow lattice thermal conductivity and high thermoelectric efficiency. Based on the first-principles calculations, we study the thermal expansion and thermomechanic properties of the experimental phase monolayer β-Te, using quasiharmonic approach. It is found β-Te shows large positive thermal expansion at elevated temperature, while the linear thermal expansion coefficient is negative along *a* direction at very low temperature. The linear thermal expansion coefficient along *b* direction is $4.9 \times 10^{-5}$ K$^{-1}$ at 500 K, which is considerably large in 2D materials. β-Te exhibits strong in-plane anisotropy, including thermal expansion, 2D elastic moduli and Poisson's ratios. However, the elastic moduli, Poisson's ratios and the in-plane

E-mail: liugang8105@gmail.com

E-mail: Xonics@tongji.edu.cn

†These authors contributed equally to this work.

anisotropy are weakened with increasing temperature, and the variations are dominated by the generalized mode Grüneisen parameters.

## INTRODUCTION

Two-dimensional (2D) materials have been investigated extensively in recent years since the exfoliation of graphene.[1-3] Due to their intriguing properties and the prospects for various device applications, the family of 2D materials has undergone rapid expansion, including transition metal dichalcogenides (TMDCs),[4-11] transition metal carbides and carbonitrides (MXenes),[12] group-III, -IV and -V monolayers.[13-20] However, there are few investigations about 2D group-VI materials so far. In 2017, 2D selenium of Group-VI has been synthesized controllably.[21] Very recently, Zhu et al.[22] predicted three phases of tellurene (Te monolayer), named α-, β-, and γ-Te. And only β-Te has been synthesized successfully on highly oriented pyrolytic graphite (HOPG) substrates by using molecular beam epitaxy.[22,23] It was found both the α- and β-Te phases possess electron and hole mobilities much higher than 2H-$MoS_2$. And this work is highlighted in an exclusive report for its potential implication.[24] Furthermore, tellurene field-effect transistors have been fabricated, showing air-stable and extraordinarily high-performance.[25] It was also found theoretically both of α- and β-Te possess excellent thermoelectric performance and ultra-low lattice thermal conductivity.[26,27,28]

Electronic devices usually work at finite temperatures. For instance, the thermoelectric efficiency of tellurene has much larger value at high temperature than room temperature.[26,27] Thus, thermal expansion and thermomechanic properties show significant importance in applications for materials. For instance, the materials will expand or contract significantly when working at high temperature, and the accumulated thermal strain and stress may affect the performance of devices greatly, or even destroy them. Thus, a more comprehensive knowledge of thermal expansion and thermomechanic properties are urgently needed and could accelerate the progress of practical applications of materials. In this work, the thermal expansion and thermomechanic properties of experimental phase β-Te are investigated by quasiharmonic approximation (QHA) based on first-principles calculations, including phonon spectra, Grüneisen parameters, thermal expansion, and temperature-dependent stiffness. It is found the linear thermal expansion coefficients (LTECs) of β-Te show significantly positive values at most temperature range studied, especially the one of $4.9\times10^{-5}$ K$^{-1}$ along *b* direction at 500 K. However, the LTEC along *a* direction is negative at low temperature. And the in-plane thermal expansion is significant anisotropic, as well as 2D elastic moduli and Poisson's ratios. At high temperature, these elastic moduli are softened, while the in-plane anisotropy becomes weaker. The variations of elastic moduli and in-plane anisotropy are also dominated by the generalized Grüneisen parameters.

## COMPUTATIONAL AND THEORETICAL METHODS

All the first-principles calculations, including the structure, electronic structure, and energy are performed using Vienna *ab initio* Simulation Package (VASP)[29-32] based on density functional theory (DFT). The exchange-correlation functional is Perdew-Burke-Ernzerhof (PBE) of generalized gradient approximation,[33,34] and the cutoff energy of the plane-wave basis is set to 500 eV. The crystal structure is relaxed with the total energy convergence criterion of $10^{-8}$ eV, while the force convergence criterion is set to $10^{-4}$ eV/Å. The effects of spin-orbit coupling (SOC) is also taken into consideration, in order to obtain accurate electron structure and other properties. During the structure optimization, the correction of long-range van der Waals (vdW) forces is taken into consideration by means of the DFT-D2 method.[35] A Monkhorst-Pack[36] k-mesh of 9×12×1 is used to sample the Brillouin zone. The phonon distributions are obtained by using the Phonopy script[37] based on the supercell approach with finite displacement method. A 4×5×1 supercell and 150×200×1 q-mesh are adopted to ensure the convergence of vibrational properties in the calculations of harmonic interatomic force constants.

In this work, quasiharmonic approximation (QHA) is adopted to investigate the temperature dependent lattice constants, which introduces the effect of temperature through the volume dependence of the vibrational frequency, and only the volume dependence is considered for the phonon anharmonicity. In the most popular scheme of QHA, the phonon spectra of about ten or more volumes are usually calculated to simulate the relationship of volume and phonon frequencies. Then the equilibrium volume at a certain temperature can be achieved through the direct minimization of the free energy. This method shows its efficiency on the investigation of thermal expansion for lots of isotropic materials, such as diamond, graphite, graphene, $h$-BN, silicene, germanene and blue phosphorene.[38-40] However, it is inefficient for the anisotropic materials. Specifically, a series of calculations needs performing over a grid of lattice-parameter points, and the dimensionality of the grid is determined by the number of independent lattice parameters.[38] It implies dozens or even hundreds of volumes of phonon spectra, which needs huge computational resources and time consumption. Besides, another shortcoming of this method is the readily emerging of negative frequency for 2D materials. It is well known that ZA mode is soft near $\Gamma$ point, and their frequency may turn negative under relatively large strain. However, to accurately fit the data points of energy to an equation of state, these data points should span in an energy range reasonably large.[39] This contradiction also affects the accuracy and validity of the method for 2D material.[39] Grüneisen theory is a method of time saving and can be used to deal with the case of anisotropic materials,[41-49] which needs the calculations of only several volumes of phonon spectra. However, in Grüneisen theory, Grüneisen parameters and elastic constants are independent of temperature and always keep constant, leading to significant deviation at high temperature probably.[50]

Within the QHA, we develop an *ab initio* method to deal with the anisotropic thermal expansion based on self-consistent quasiharmonic approximation (SC-QHA).[51-54] It can investigate thermal expansion not only with high accuracy but also time saving. However, it does not adopt the scheme of self-consistent iteration like SC-QHA, but it achieves equilibrium lattice constants through pressure solving directly, named *pressure-solving quasiharmonic approximation* (PS-QHA). Furthermore, we generalize it to the case of anisotropic thermal expansion.

The total free energy consisting of electronic energy $E(a_i)$ and phonon free energy $F_{ph}(a_i)$, can be expressed as:

$$F_{tot} = E(a_i) + F_{ph}(a_i) = E(a_i) + \frac{1}{N}\sum_{\lambda,\mathbf{q}}\left\{\frac{1}{2}\hbar\omega_{\lambda,\mathbf{q}}(a_i) + k_B T \ln[1-\exp(-\frac{\hbar\omega_{\lambda,\mathbf{q}}(a_i)}{k_B T})]\right\}, \quad (1)$$

where $a_i$ means the independent lattice parameters, i.e., $a$ and $b$ for β-Te. $k_B$, $\hbar$ and $N$ are the absolute temperature, Boltzmann constant, reduced Planck constant, and the number of q-point in BZ, respectively. $E(a_i)$ is the ground-state free energy, $\omega_{\lambda,\mathbf{q}}$ is the phonon frequency corresponding to wave vector $\mathbf{q}$, mode $\lambda$. In the QHA, as $\omega_{\lambda,\mathbf{q}}$ is only volume-dependent, $\omega_{\lambda,\mathbf{q}}$ and generalized mode Grüneisen parameter $\gamma_{\lambda,\mathbf{q}}(a_i)$ can be described by a Taylor expansion, up to the second order:

$$\omega_{\lambda,\mathbf{q}}(a_i) = \omega_{\lambda,\mathbf{q}}(a_{i,0}) + \sum_i \left(\frac{\partial \omega_{\lambda,\mathbf{q}}}{\partial a_i}\right)_0 \Delta a_i + \frac{1}{2}\sum_{i,j}\left(\frac{\partial^2 \omega_{\lambda,\mathbf{q}}}{\partial a_i \partial a_j}\right)_0 \Delta a_i \Delta a_j, \quad (2)$$

$$\gamma_{\lambda,\mathbf{q}}(a_i) = -\frac{a_i}{\omega_{\lambda,\mathbf{q}}} \cdot \left[\left(\frac{\partial \omega_{\lambda,\mathbf{q}}}{\partial a_i}\right)_0 + \left(\frac{\partial^2 \omega_{\lambda,\mathbf{q}}}{\partial a_i^2}\right)_0 \Delta a_i\right]. \quad (3)$$

Here $a_{i,0}$ is the reference lattice constant which is obtained directly by the geometry optimization of DFT, and $\Delta a_i = a_i - a_{i,0}$. The equilibrium state under the zero external pressure $P$, fulfills the relationship:

$$\frac{\partial F_{tot}}{\partial a_i} = \frac{\partial F_{tot}}{\partial V} \cdot \frac{\partial V}{\partial a_i} = P\frac{\partial V}{\partial a_i} = 0, \quad (4)$$

In fact, $\frac{\partial F_{tot}}{\partial a_i}$ can be considered as the force acts on the area perpendicular to $a_i$

direction, and it also can be expressed as:

$$\frac{\partial F_{tot}}{\partial a_i} = \frac{\partial E}{\partial a_i} + \frac{\partial F_{ph}}{\partial a_i}$$

$$= \frac{\partial E}{\partial a_i} - \frac{\hbar}{Na_i} \sum_{\lambda,\mathbf{q}} \omega_{\lambda,\mathbf{q}}(a_i) \cdot \gamma_{\lambda,\mathbf{q}}(a_i) \cdot \left\{ \frac{1}{2} + \frac{1}{\left[\exp\left(\frac{\hbar \omega_{\lambda,\mathbf{q}}(a_i)}{k_B T}\right) - 1\right]} \right\} = 0 \quad (5)$$

Then we can obtain the equilibrium lattice constants at a certain temperature by solving Eqs. (2), (3) and (5). Actually, when the external pressure is zero, the equilibrium lattice constants are determined by the balance of $\frac{\partial E}{\partial a_i}$ and $\frac{\partial F_{ph}}{\partial a_i}$, i.e. the forces contributed by electron and phonon respectively. At last, the linear thermal expansion coefficient (LTEC) can be achieved through:

$$\alpha_i = \frac{1}{a_i} \frac{\partial a_i}{\partial T}. \quad (6)$$

Moreover, the temperature dependent 2D elastic constant can be obtained based on the total free energy, through following equation:

$$C_{ij} = \frac{1}{S} \cdot \frac{\partial^2 F_{tot}}{\partial \epsilon_i \partial \epsilon_j} = \frac{1}{S} \cdot \left( \frac{\partial^2 E}{\partial \epsilon_i \partial \epsilon_j} + \frac{\partial^2 F_{ph}}{\partial \epsilon_i \partial \epsilon_j} \right), \quad (7)$$

where $S$ is the unit area, $\epsilon_i$ means strain tensor, and the Voigt notation is adopted. Note the first term of right side is the contribution of electron to 2D elastic constant, while the second term means the contribution of phonon, which can be named as electronic and phonon elastic constant. The latter can be expressed as:

$$\frac{1}{S}\cdot\frac{\partial^2 F_{ph}}{\partial \epsilon_i \partial \epsilon_j} = \frac{a_i a_j}{S} \cdot \frac{\hbar}{N} \sum_{\lambda,\mathbf{q}} \left\{ \left( \frac{1}{2} + \frac{1}{\exp\left(\frac{\hbar\omega_{\lambda,\mathbf{q}}}{k_B T}\right)-1} \right) \cdot \frac{\partial^2 \omega_{\lambda,\mathbf{q}}}{\partial a_i \partial a_j} \right.$$
$$\left. - \frac{\exp\left(\frac{\hbar\omega_{\lambda,\mathbf{q}}}{k_B T}\right)}{\left[\exp\left(\frac{\hbar\omega_{\lambda,\mathbf{q}}}{k_B T}\right)-1\right]^2} \cdot \frac{\hbar}{k_B T} \cdot \frac{\partial \omega_{\lambda,\mathbf{q}}}{\partial a_i} \cdot \frac{\partial \omega_{\lambda,\mathbf{q}}}{\partial a_j} \right\} . \quad (8)$$

In this work, first, electronic energies of 48 sets of lattice constants around the equilibrium lattice constants are calculated using DFT. Specifically, $a/a_0$ are chosen from 0.98 to 1.03, and $b/b_0$ are in the range of 0.98 to 1.05. Both of them are with the step of 0.01. These electronic energies are interpolated with cubic spline function, to obtain $\frac{\partial E}{\partial a_i}$. Then five sets of lattice constants of β-Te are considered for the calculation of phonon spectra: $(0.99a_0, 1.00b_0)$, $(1.00a_0, 1.00b_0)$, $(1.01a_0, 1.00b_0)$, $(1.00a_0, 0.99b_0)$, and $(1.00a_0, 1.01b_0)$. Note $a_0$ and $b_0$ denote the reference lattice constants obtained by DFT directly. The internal coordinates of the atoms for each structure under strains are relaxed to include the effect of the displacements of them on the phonon property.[45] Usually, it is sufficient for the materials without phase transition. The total five sets of phonon spectra are calculated, to obtain the first and second order derivatives of phonons. Then phonon frequency and the generalized Grüneisen parameter for any set of $(a, b)$ can be also obtained based on Eq. (2) and (3). Furthermore, total free energy at a certain temperature $T$ can be determined based on Eq. (1). After that, we can solve Eq. (5) by using dichotomy. Note there are two independent variables of $a$ and $b$. We firstly fix $b = b_0$, then perform dichotomy method for $a$, until $\frac{\partial F_{tot}}{\partial a} < eps$, here $eps$ is the stopping criterion. Now we denote the convergent $a$ from dichotomy as $a_n$. Next, with fixed $a = a_n$, we perform dichotomy method for $b$, until $\frac{\partial F_{tot}}{\partial b} < eps$, and denote final $b$ as $b_n$. Then we can also check whether $\frac{\partial F_{tot}}{\partial a} < eps$ is satisfied. If the two conditions are satisfied simultaneously, the calculation finishes. Otherwise, we should repeat above procedures, until the two conditions are satisfied simultaneously. The last $a_n$ and $b_n$ are the solution. After the lattice constants are calculated at any $T$, TECs and the temperature dependent elastic constants can be calculated based on Eq. (6), (7) and (8). Compared with the conventional QHA of minimization of free energy, our scheme can not only save lots of computational time for phonon spectra, but also avoid the negative frequency for 2D materials when large strain applied.

## RESULTS AND DISCUSSION

First of all, we focus on the calculation of LTECs of in-plane isotropic 2D materials to test the validity and reliability of our new PS-QHA. The LTECs of 2H-MoS$_2$ and 2H-MoSe$_2$ are calculated and shown in Fig. 1. It is found the lines of PS-QHA are very closed to the results of previous works,[52,54] indicating our PS-QHA is accurate and reliable.

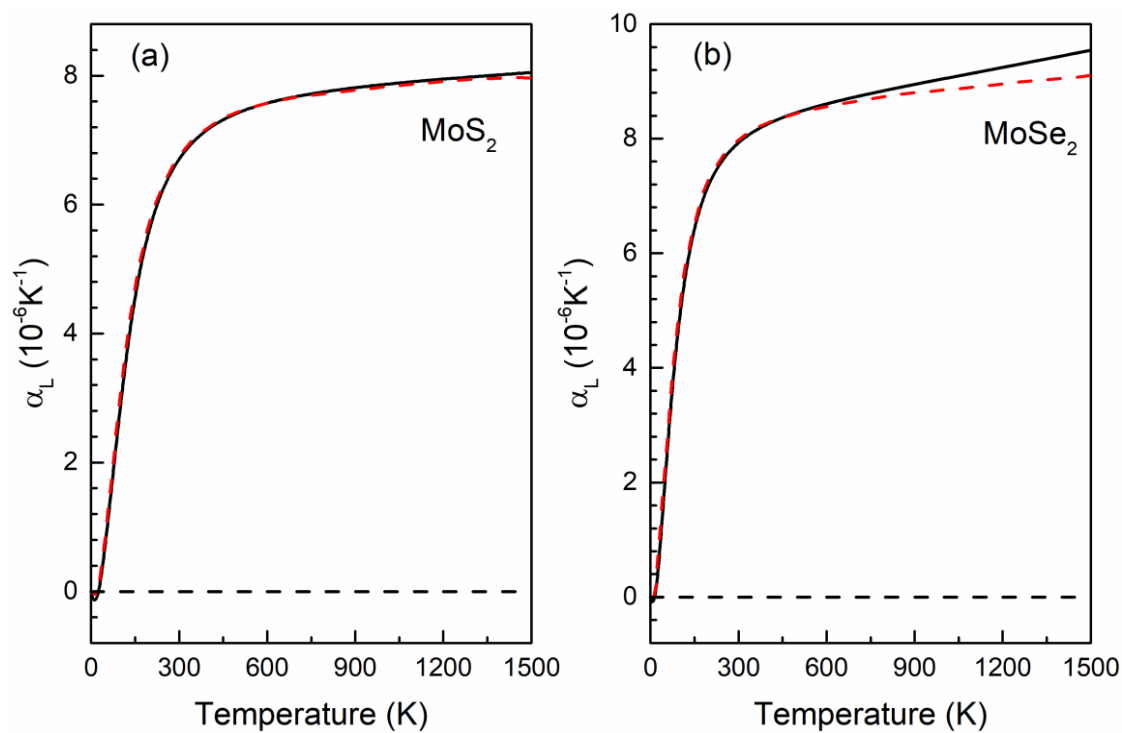

Fig. 1. The LTECs of 2H- H-MoS$_2$ and 2H-MoSe$_2$, compared with previous *ab initio* calculations. The black solid lines in (a) and (b) present the results of PS-QHA, and the red dashed lines are from the results of Ref. 52 and Ref. 54

The optimized structures of β-Te are displayed in Fig. 2(a). The optimized lattice parameters are $a$ = 5.499 and $b$ = 4.173 Å. The buckling height $h$ = 2.176 Å. The data we obtained are in good agreement with previous studies.[22,23,26-28] Note here $h$ is much thicker than many other 2D materials, such as silicene of 0.42~0.45 Å,[39,53] germanene of about 0.69 Å,[39,53] blue phosphorene of 1.24 Å.[39] Actually, it can be compared to that of 2.51 Å for black phosphorene.[55] It is worth noting that the relatively large buckling height affects much on atom vibrations and generalized mode Grüneisen parameter, which will be discussed later.

A detailed knowledge of the phonon dispersions is a prerequisite not only for the performance of QHA, but also for the understanding of various phononic and thermomechanic properties. The calculated phonon dispersions of equilibrium lattice constants, as well as the ones under the strains we chosen, are shown in Fig. 2(b) and (c). The projected density of states (PDOS) for the equilibrium structure is also exhibited in Fig. 2(d). It shows that all phonon branches of equilibrium lattice constants are positive without negative frequency, confirming the dynamical stability of β-Te.[28] The range of phonon dispersion is 0 to about 5.5 THz. The out-of-plane (Z) and in-plane (X and Y) vibrations couple with each other in the whole frequency range, different from graphene.[56] The in-plane C-C bonds are orthogonal to the out-of-plane direction in planar graphene, resulting in the complete decoupling of in-plane and out-of-plane vibrations.[52,53,56] However, in β-Te with large thickness, the covalent bonds become nonorthogonal, leading to the hybridization of these vibrations. Furthermore, the phonon under tensile or compressed strains along two directions, which are calculated by Eq. (2), are free of imaginary frequency, too. It ensures the validity of calculations of thermal expansion. It is found the variations of phonon spectra along $a$ direction are smaller than the ones along $b$ direction, indicating the smaller generalized mode Grüneisen parameters along $a$ direction.

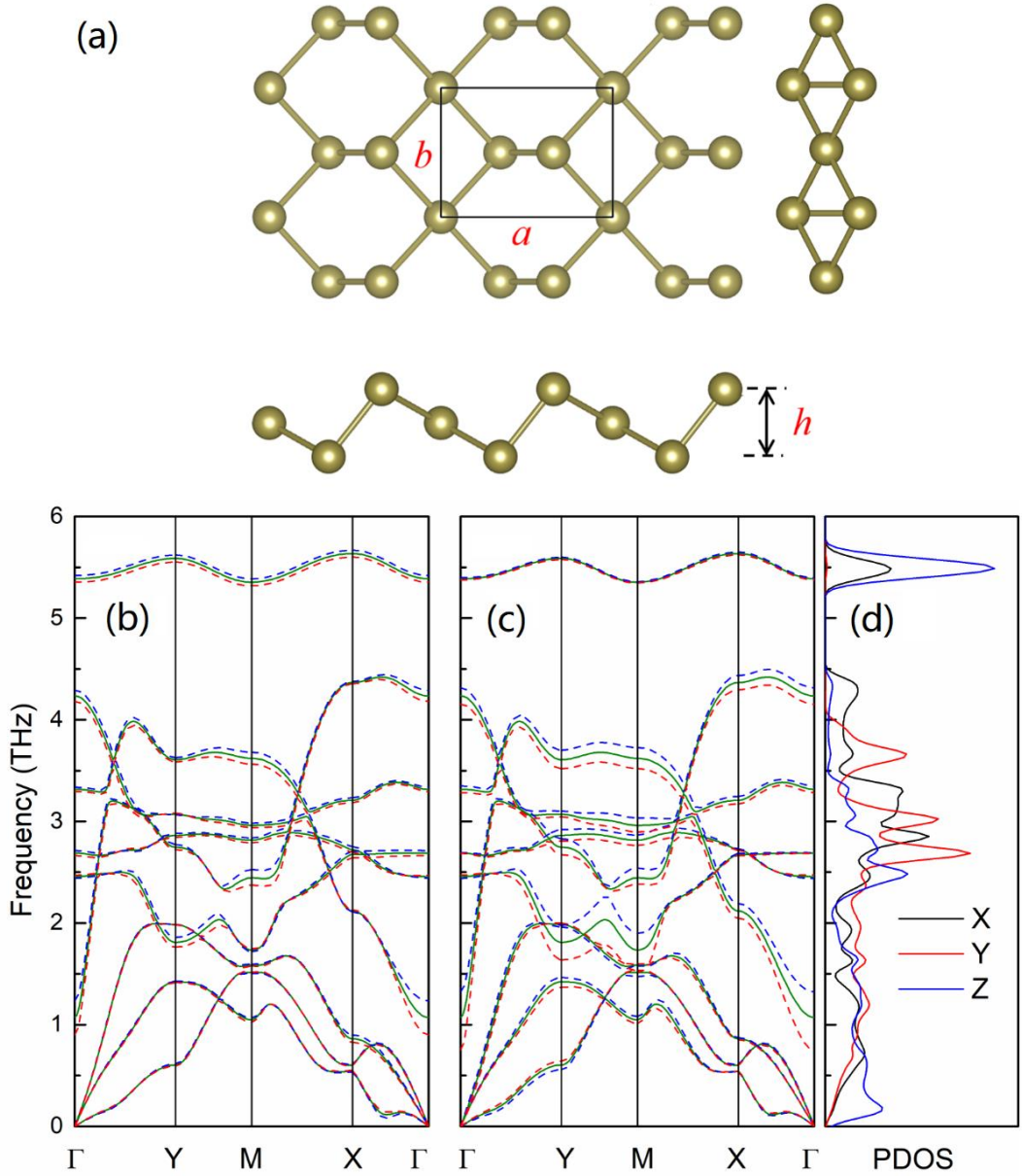

Fig. 2. (a) The optimized structures of β-Te. Both top view and side views are exhibited, and primitive cell is marked with black solid lines. Note *h* is buckling height. (b) and (c) show Phonon spectra of β-Te along the high symmetric line under strains, and green solid lines always represent phonon dispersion free of strain. Blue and red dashed lines mean the phonon spectra under ±1% strain along *a* direction in (b), while they represent the ones under ±1% strain along *b* direction in (c). The Projected DOS of equilibrium lattices along X, Y and Z directions are displayed in (d).

As an important physical parameter closely related phonon anharmonicity, generalized mode Grüneisen parameter *γ* is investigated by applying the strains of ±1% along *a* and *b* directions respectively, as exhibited in Fig. 3. It should be noted that *γ* of most

ZA phonons have large negative values around Γ, as displayed in Fig. 3. It is in agreement of the well-known phenomenon that ZA phonons around Γ show large negative $\gamma$ for 2D material, such as graphene,[40,42] h-BN,[40] and MoS$_2$.[40,52] It is named as membrane effect:[38] in a 2D system, the vibrations of bending ZA modes can be compared to the ones in a string. When the string is stretched, it will be stiffer with vibrations of smaller amplitude but higher frequency, leading to negative $\gamma$. However, there are also many ZA phonons have positive $\gamma$, due to its relatively large thickness. This is different from the case of planar graphene and h-BN, whose ZA phonons clearly show total negative value of mode Grüneisen parameters.[40,42] It is similar to MoS$_2$, whose structure is nonplanar.[40,52] It originates from larger thickness introduces more hybridization of the in-plane and out-of-plane vibrations, leads to softening effect on the ZA modes in thicker membrane. It is so called thickness effect,[57] which can counteract the membrane effect in 2D systems. And it can significantly affect its thermal expansion at low temperature, which will be discussed later in our work. Furthermore, the distribution range of most $\gamma(a)$ are significantly smaller than that of $\gamma(b)$, showing remarkable anisotropy. The unique structure of β-Te along $a$ direction, which is similar to the black phosphorene along armchair direction to some extent, determines β-Te is softer along $a$ direction.[58] Thus, the tension along $a$ direction can change the bond properties less than that along $b$ direction, indicating smaller variation of phonon frequencies and $\gamma$. The optical phonons around 1 THz have the greatest $\gamma$, implying relatively larger phonon anharmonicity, compared with other optical phonons. And after careful examination, we find these optical phonons are corresponding to the vibrations along $y$ direction ($b$ direction), in consistent with previous study.[28] In fact, the anharmonic frozen-phonon potential curve of this phonon mode in Ref. 28 illustrates the relatively large anharmonicity, which can be reproduced excellently by a fourth order polynomial fitting, while the second order polynomial fitting shows significant deviation. It also implies that the large $\gamma$ originates from the nonlinear dependence of restoring forces on atomic displacement amplitudes, which is also the direct evidence

of the anharmonicity.[28,59,60]

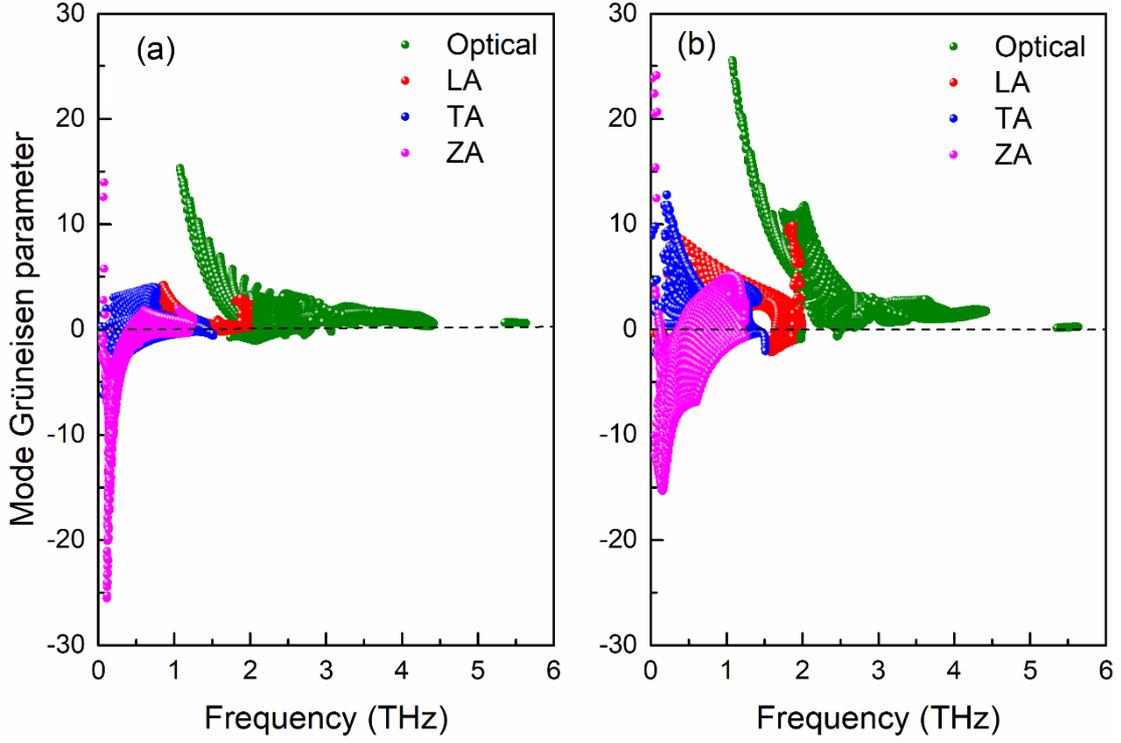

Fig. 3. The generalized mode Grüneisen parameters $\gamma$. (a) shows $\gamma(a)$ corresponding the strain along $a$ direction, and $\gamma(b)$ corresponding the strain along $b$ direction are exhibited in (b).

Macroscopic Grüneisen parameters can be defined as:[61,62]

$$\bar{\gamma}(a_i) = \frac{\sum_{\lambda,\mathbf{q}} C_{\lambda,\mathbf{q}} \gamma_{\lambda,\mathbf{q}}(a_i)}{\sum_{\lambda,\mathbf{q}} C_{\lambda,\mathbf{q}}}, \qquad (9)$$

where $C_{\lambda,\mathbf{q}}$ is the mode contributions to the heat capacity. It represents the average value of the generalized mode Grüneisen parameter weighted by $C_{\lambda,\mathbf{q}}$. Here $\bar{\gamma}$ are calculated and shown in Fig. 4. Note based on Eq. (3), generalized mode Grüneisen parameter $\gamma$ changes with lattice constants, which depends on temperature. It can be found that $\bar{\gamma}$ along both directions show the similar dependence of temperature. They are all significant negative values at very low temperature, as the ZA phonons with low frequency and negative $\gamma$ can be activated readily. Then they rise quickly and become positive. The corresponding critical temperatures are 20 and 10 K for $\bar{\gamma}(a)$ and $\bar{\gamma}(b)$.

At last they reach their saturation values. However, the saturation value of $\bar{\gamma}(b)$ is 1.61, remarkably larger than 0.65 of $\bar{\gamma}(a)$. This is consistent with the results of mode Grüneisen parameter $\gamma$, as shown in Fig. 3. The macroscopic Grüneisen parameters $\bar{\gamma}_0$, based on the mode Grüneisen parameter $\gamma_0$ corresponding to the reference lattice constants $a_0$ and $b_0$, are also displayed for comparison. In general, $\bar{\gamma}$ and $\bar{\gamma}_0$ are very closed to each other, especially the those along $a$ direction. At high temperature, there becomes small difference between the lines of $\bar{\gamma}(b)$ and $\bar{\gamma}_0(b)$. It implies $\gamma$ change little with the change of lattice constants and are independent of temperature approximatively.

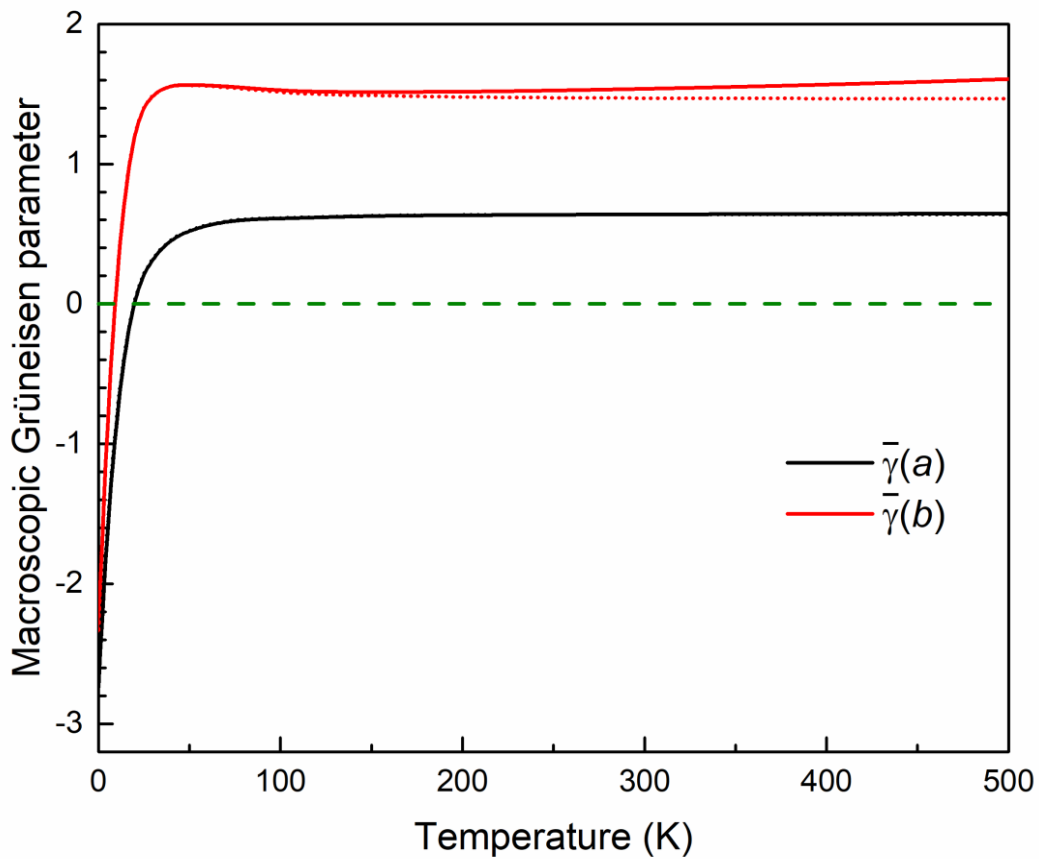

Fig. 4. Macroscopic Grüneisen parameters $\bar{\gamma}$ vary with temperature. Short dotted lines indicate $\bar{\gamma}_0$, corresponding to the reference lattice constants $a_0$ and $b_0$. Note the line of $\bar{\gamma}_0(a)$ nearly overlaps the one of $\bar{\gamma}(a)$.

The ratios of lattice constants $a(T)/a_0$ and $b(T)/b_0$ are displayed in Fig. 5(a). Firstly, the lattice constants at 0 K i.e. $a(0)$ and $b(0)$, are not equal to the reference lattice constants $a_0$ and $b_0$. They expand by 0.19% and 0.08% along $a$ and $b$ directions, respectively. It originates from the contribution of zero-point vibration, which is also included in the most popular scheme of QHA, whereas it is omitted in Grüneisen theory. The lattice constants $b$ stretches greatly with the increasing temperature. And it shows remarkable anisotropy. At 500 K, $b$ expands by about 1.6% while $a$ expands by only 0.7%.

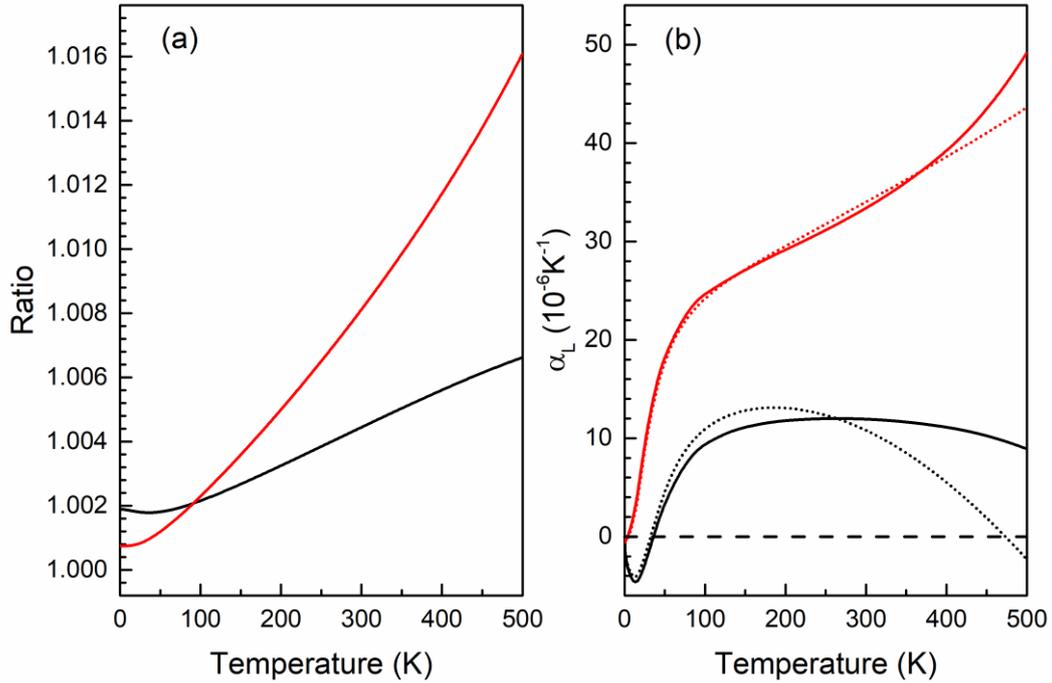

Fig. 5. The ratios of lattice constants (a) and TECs (b) for β-Te. The black and red lines represent the data along $a$ and $b$ directions respectively. And the short dotted lines mean the TECs from conventional QHA in (b).

We also plot the temperature dependence of LTECs in Fig. 5(b). It is found β-Te shows giant thermal expansion, especially along $b$ direction. At 500 K, the LTEC along this direction $\alpha_2$ is as large as $4.9\times 10^{-5}$ K$^{-1}$, much larger than many 2D materials, such as black phosphorene ($9\sim 10\times 10^{-6}$ K$^{-1}$),[58,63] $MoS_2$ ($7.5\times 10^{-6}$ K$^{-1}$),[52] $MoSe_2$ ($8\times 10^{-6}$ K$^{-1}$),[54] and $MoTe_2$ ($4\times 10^{-6}$ K$^{-1}$).[54] The LTEC along $a$ direction $\alpha_1$ is $8.9\times 10^{-6}$ K$^{-1}$, much smaller than the one along $b$ direction at the same temperature. In fact, at 500 K, the ratio of the LTEC along $a$ direction to that along $b$ direction is about 0.18, indicating giant anisotropic thermal expansion. For comparison, the ratio of the LTECs is about 0.39 for black phosphorene at high temperature.[58] It is concluded that β-Te shows much stronger in-plane anisotropy of thermal expansion than black phosphorene. Below 35 K, $α_1$ is negative, and it reaches the lowest value of -4.6×10$^{-6}$ K$^{-1}$ at 15 K. It is similar to the case of MoS$_2$, which also has small negative thermal expansion only in a narrow temperature range.[40,52] On the contrary, graphene and *h*-BN show significantly negative thermal expansion from 0 K to high temperature.[40] The difference originates from their planar or nonplanar structures, which affect mode Grüneisen parameters around Γ point. The most negative value of Mode Grüneisen parameters of ZA phonons is about -25 in β-Te, much higher than those of planar graphene and *h*-BN (as low as -80),[40] indicating there can be only a small temperature range of negative thermal expansion in β-Te. On the other side, $α_2$ is positive in the whole temperature range. It is noteworthy that $α_2$ is positive while $\bar{γ}(b)$ are negative in the range of 0 to 10 K, contrary to the common perspective for isotropic material, that negative macroscopic Grüneisen parameter usually leads to negative thermal expansion,[38,39] and vice versa. It can be understood using Grüneisen theory:[34]

$$\begin{cases} α_1 = \dfrac{C_V}{A_0}(S_{11}\bar{γ}(a) + S_{12}\bar{γ}(b)) \\ \\ α_2 = \dfrac{C_V}{A_0}(S_{12}\bar{γ}(a) + S_{22}\bar{γ}(b)) \end{cases} \quad (10)$$

Here $A_0$ is the area of primitive cell, and $S_{ij}$ is matrix element of elastic compliance tensor, which is the inverse of the elastic stiffness tensor:

$$S_{11} = \dfrac{C_{22}}{C_{11}C_{22} - C_{12}^2}, \quad S_{22} = \dfrac{C_{11}}{C_{11}C_{22} - C_{12}^2}, \quad S_{12} = -\dfrac{C_{12}}{C_{11}C_{22} - C_{12}^2}. \quad (11)$$

Based on Eq. (10), it is found macroscopic Grüneisen parameter can determine the sign of LTEC as $\bar{γ}(a) = \bar{γ}(b)$ in isotropic materials. However, macroscopic Grüneisen parameter can't solely determine the sign of LTEC for anisotropic materials. For instance, $\bar{\gamma}(a)$, $\bar{\gamma}(b)$ are -1.636, -0.881 respectively, and $S_{11}$, $S_{12}$, $S_{22}$ are 0.095, -0.025, 0.043 m N$^{-1}$ at 5 K. This combination determines LTEC is positive though $\bar{\gamma}(b)$ is negative.

To confirm the validity and reliability of our PS-QHA, we also calculate TECs for β-Te by conventional QHA. Here we construct a grid of 54 points of lattice constants. Specifically, $a/a_0$ is chosen from 0.99 to 1.015, and $b/b_0$ is in the range of 0.99 to 1.03, with both steps of 0.005. The results are also plotted by short dotted lines in Fig. 5(b) for comparison. It can be found the data of two methods are in reasonable agreement with each other, especially at low temperature. However, at high temperature, there can be found significant deviations. After carefully checking, we find the remarkable deviations come from the inevitable imaginary frequency in conventional QHA. In conventional QHA, a 2D grid is constructed, where the phonon spectra of each lattice parameter point need calculating. The points used by conventional QHA and those β-Te can actually reach during thermal expansion, as well as the ones which are corresponding to imaginary frequencies, are displayed in Fig. 6(a). It should be noted the grid is for the fitting of the curve surface of total free energies, though most points in the grid are not corresponding to the points which the cell of β-Te can actually reach in the process of thermal expansion, as shown in Fig. 6(a). Moreover, the points β-Te can actually reach at high temperature are closed to the region of imaginary frequency. It causes that these additional points can't be removed in the conventional QHA method, though they correspond to imaginary frequency. It is because in conventional QHA, the range of grid should be large enough to ensure the accuracy and reliability of the fitting result. This contradiction causes that the result of conventional QHA can't be quite reliable for β-Te. However, our PS-QHA can circumvent the problem of imaginary frequencies, as displayed in Fig. 6(a). The phonon spectra of β-Te at 300 and 500 K are plotted Fig. 6(b), without any imaginary frequency, indicating the validity and reliability of our PS-QHA. There is no structure with negative frequency involved in PS-QHA, indicating our new method is more reliable and accurate than conventional

QHA for β-Te. In fact, previous work[39] has already pointed out that conventional QHA is not very applicable for 2D materials with large thermal expansion, as imaginary frequencies probably occur under large strain in 2D materials.

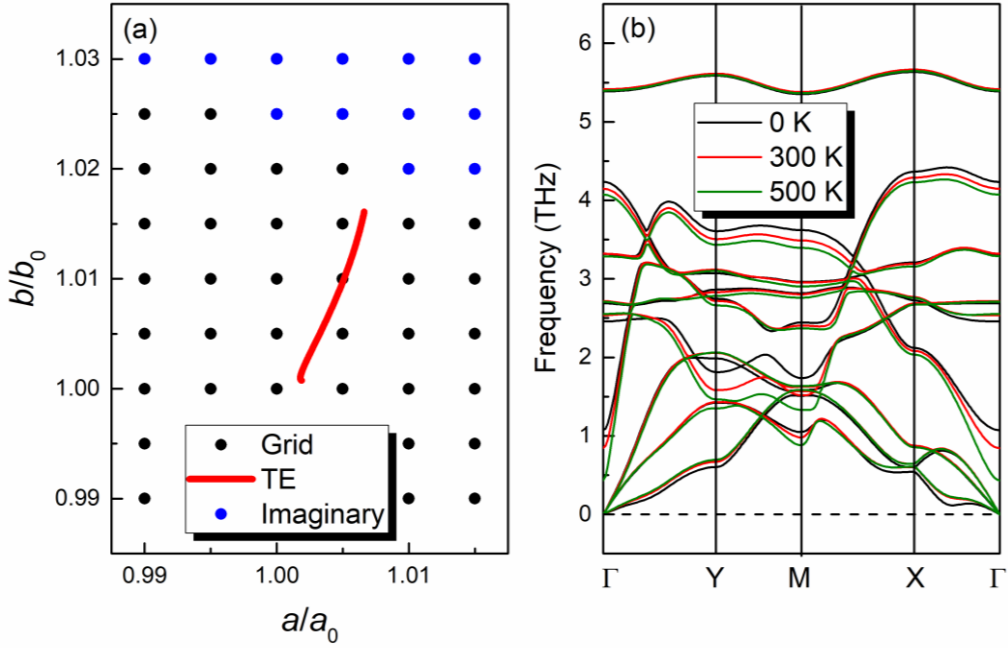

Fig. 6. (a) shows the points of lattice parameters used in conventional QHA, displayed by black solid circles. The red line shows the structures β-Te can actually reach during thermal expansion. Blue solid circles mean the points corresponding to imaginary frequencies. (b) exhibits the phonon spectra of β-Te at 0, 300, and 500 K, respectively. Note the lattice constants of β-Te at 300 and 500 K are $(1.004a_0, 1.008b_0)$ and $(1.007a_0, 1.016b_0)$, respectively.

Based on the temperature dependent elastic constants, we can obtain the temperature dependent 2D bulk moduli, Young's moduli, and Poisson's ratios:[55,64-67]

$$B_{2D} = \frac{1}{4}(C_{11} + C_{22} + 2C_{12}),$$
$$E_x = \frac{C_{11}C_{22} - C_{12}C_{21}}{C_{22}}, \quad E_y = \frac{C_{11}C_{22} - C_{12}C_{21}}{C_{11}}, \quad (12)$$
$$v_{xy} = \frac{C_{21}}{C_{22}}, \quad v_{yx} = \frac{C_{12}}{C_{11}}.$$

Note $C_{12}$ is equal to $C_{21}$, and the unit of these 2D elastic moduli is N m$^{-1}$, which can be converted to bulk unit of N m$^{-2}$ by dividing the effective thickness of material. These elastic moduli and Poisson's ratios are calculated and plotted in Fig. 7, and show significant in-plane anisotropy. The elastic constant and bulk modulus along *a* direction ($C_{11}$ and $E_x$) are much smaller than those along *b* direction ($C_{22}$ and $E_y$), implying β-Te is much softer along *a* direction. This is also in agreement with the distribution of generalized mode Grüneisen parameters. Moreover, $C_{22}$ and $E_y$ are more sensitive to temperature than $C_{11}$ and $E_x$. From 0 to 500 K, elastic moduli reduce by 15%, 35%, 17% and 36% for $C_{11}$, $C_{22}$, $E_x$ and $E_y$, respectively. All the elastic moduli decrease with increasing temperature, indicating the softening of β-Te at high temperature. Based on Eq. (12), the Poisson's ratios change with the variations of elastic constants $C_{11}$, $C_{22}$, and $C_{12}$ at elevated temperature. It is notable the sign of Poisson's ratio doesn't have direct relation to the sign of thermal expansion coefficients, as positive/negative thermal expansion is not solely determined by elastic constants based on Grüneisen theory. These elastic constants are always positive, determining the positive Poisson's ratios in the temperature range we studied. The Poisson's ratio $v_{xy}$ rises while $v_{yx}$ declines when temperature becomes higher. The different temperature dependent behaviors originate from the faster decline of $C_{22}$ than $C_{12}$ with increasing temperature, whereas $C_{12}$ decreases faster than $C_{11}$. All these results above show the in-plane anisotropy is weakened by increasing temperature.

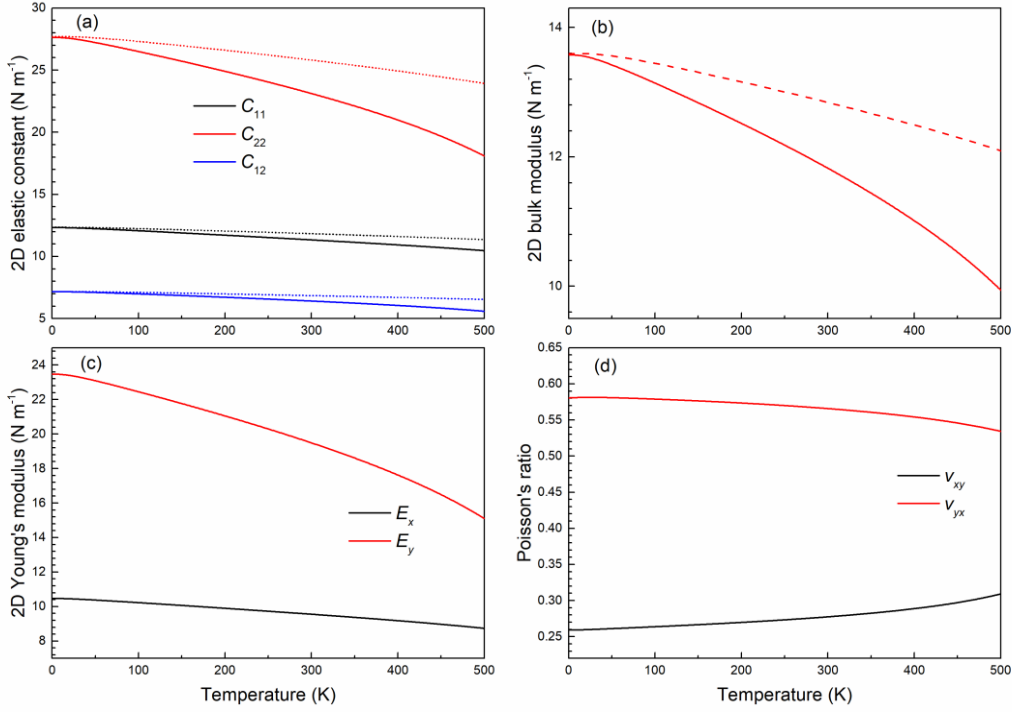

Fig. 7. The temperature dependences of 2D elastic constants (a), bulk moduli (b), Young's moduli (c), and Poisson's ratios (d). Note the short-dotted lines represent the physical quantities considering the contribution of electron only.

Since the total free energy consists of electronic energy and phonon free energy, the contributions of electron to elastic constants and bulk modulus (electronic elastic constants and bulk modulus) are also displayed with short-dotted lines for comparison, as shown in Fig. 7. It is assumed that the temperature dependences of elastic moduli are solely caused by lattice thermal expansion, without the contribution of phonons. The electronic elastic moduli also reduce with increasing temperature, as volume expansion always causes the decease of the second derivative of the potential-energy surface.[57] The electronic elastic moduli are nearly the same with the total elastic moduli at 0 K, as the contribution of phonon trends toward zero at this temperature, based on Eq. (8). In fact, at 0 K the phonon elastic constants can be expressed as:

$$\lim_{T \to 0} \frac{1}{S} \cdot \frac{\partial^2 F_{ph}}{\partial \epsilon_i \partial \epsilon_j} = \frac{a_i a_j}{S} \cdot \frac{\hbar}{N} \sum_{\lambda,\mathbf{q}} \frac{1}{2} \cdot \frac{\partial^2 \omega_{\lambda,\mathbf{q}}}{\partial a_i \partial a_j}. \qquad (13)$$

Thus, it can be concluded that the second order derivative of $\omega$ are very small and can be ignored safely, since the electronic elastic constants are equal to the total elastic constant at 0 K, indicating the phonon elastic constants trend to zero. Moreover, it is in agreement of Fig. 4, which implies mode Grüneisen parameter $\gamma$ doesn't vary much along with lattice constants. At high temperature, there are significant derivations between electronic and total elastic moduli, which originates from the intense excitation of phonons at high temperature. The reductions of electronic elastic moduli only represent minor percentages of the whole reductions. For instance, the reduction of electronic bulk modulus is only 41% of the reduction of total bulk modulus in the temperature range. It indicates phonon dominates the temperature dependences of elastic moduli in β-Te, the same as many other 2D materials.[50] The effect of phonon must be taken into account when the elastic properties are investigated at elevated temperature. Based on Eq. (8), at high temperature the contribution of phonon to elastic constants can be expressed approximately as:

$$\lim_{T\to\infty}\frac{1}{S}\cdot\frac{\partial^2 F_{ph}}{\partial\epsilon_i\partial\epsilon_j}=-\frac{1}{S}\cdot\frac{k_B T}{N}\sum_{\lambda,\mathbf{q}}\left(\gamma_{\lambda,\mathbf{q}}(a_i)\cdot\gamma_{\lambda,\mathbf{q}}(a_j)\right). \quad (14)$$

Note the second derivative of frequency $\omega$ is ignored here. It is obvious that generalized mode Grüneisen parameters have an important effect on the temperature dependence of elastic moduli. On the whole, $\gamma_{\lambda,\mathbf{q}}(a)$ are much smaller than $\gamma_{\lambda,\mathbf{q}}(b)$, leading to the smaller reduction of $C_{11}$ than that of $C_{22}$ at high temperature, as shown in Fig. 7. It also determines the faster decline of $C_{22}$ than $C_{12}$, as well as more rapidly decrease of $C_{12}$ than $C_{11}$. Furthermore, it also indicates that the temperature dependence of elastic constant is nonlinear, as mode Grüneisen parameters vary with lattice constants, which also change with temperature. It is in agreement with Fig. 7(a). It is concluded reasonably that the generalized mode Grüneisen parameters also play an important role in the variations of elastic moduli, Poisson's ratios and in-plane anisotropy along with temperature.

## CONCLUSIONS

In summary, we investigate the anisotropic thermal expansion and thermomechanic namic properties for β-Te using first-principles calculations. Based on QHA, we develop a new scheme named PS-QHA to study anisotropic thermal expansion of β-Te with high accuracy and time saving, which needs only five sets of phonon spectra. β-Te shows giant positive thermal expansion along $b$ direction, which is about $4.9\times10^{-5}$ $K^{-1}$ at 500 K. The in-plane thermal expansion is significantly anisotropic. The LTEC along $a$ direction is about 18% of that along $b$ direction at 500 K. Phonon spectra, generalized mode Grüneisen parameters $\gamma$ and macroscopic Grüneisen parameters $\bar{\gamma}$ are also studied. The values of $\gamma(a)$ are found disperse within a narrower range than those of $\gamma(b)$, due to the material is softer along $a$ than $b$ direction. And it leads to smaller saturation value of $\bar{\gamma}(a)$ than $\bar{\gamma}(b)$ at high temperature. The comparison of PS-QHA and conventional QHA is displayed, with the conclusion that PS-QHA is more reliable and valid than conventional QHA for β-Te, while this conclusion can be extended to 2D materials with large thermal expansion. Furthermore, the temperature dependent 2D elastic constants, bulk modulus, Young's modulus and Poisson's ratios are exhibited. All of them show intense in-plane anisotropy, while it becomes weaker at elevated temperature. The electronic contributions to the variations of these 2D elastic moduli with increasing temperature are not dominant. Phonon contributes most to the variations and can't be ignored in the investigation of thermomechanic properties at high temperature, while the variations of elastic moduli, Poisson's ratios and in-plane anisotropy are also dominated by the generalized mode Grüneisen parameters. Our work is likely to be of value not only for the potential applications of β-Te such as a thermoelectric material, but also for the development of theoretical research of anisotropic thermal expansion. Our new scheme can be expected to investigate anisotropic thermal expansion of other materials efficiently.

## ACKNOWLEDGMENTS

This work is supported by the National Natural Science Foundation of China (No. 11775159), the Natural Science Foundation of Shanghai (No. 18ZR1442800), the